\documentclass[prd,twocolumn,showpacs,amsmath,amssymb,floatfix]{revtex4}
\usepackage{graphicx,color,dcolumn,booktabs,bm}
\usepackage{longtable,lscape}
\usepackage{txfonts}
\usepackage{amssymb}
\usepackage{indentfirst}

\begin{document}
\title{The observed charmed hadron $\Lambda_c(2940)^+$ and the $D^*N$ interaction}
\author{Jun He$^{1,3}$}
\author{Xiang Liu$^{1,2}$\footnote{corresponding author}}\email{xiangliu@lzu.edu.cn}
\affiliation{
$^1$Research Center for Hadron and CSR Physics,
Lanzhou University $\&$ Institute of Modern Physics of CAS, Lanzhou 730000, China\\
$^2$School of Physical Science and Technology, Lanzhou University, Lanzhou 730000,  China\\
$^3$Institute of Modern Physics of CAS, Lanzhou 730000, China}
\date{\today}


\begin{abstract}

In this work, we systematically study the interaction of $D^*$ and
nucleon, which is stimulated by the observation of
$\Lambda_c(2940)^+$ close to the threshold of $D^*p$. Our numerical
result obtained by the dynamical investigation indicates the existence
of the $D^*N$ systems with $J^P=\frac{1}{2}^\pm,\,\frac{3}{2}^\pm$, which not
only provides valuable information to understand the underlying
structure of $\Lambda_c(2940)^+$ but also improves our knowledge of
the interaction of $D^*$ and nucleon. Additionally, the bottom
partners of the $D^*N$ systems are predicted, which might be as one of the tasks in LHCb experiment.

\end{abstract}

 \maketitle

\section{Introduction}\label{sec1}

The BaBar Collaboration reported a new charmed hadron $\Lambda_c(2940)^+$ with mass $M=2939.8\pm1.3(\mathrm{stat})\pm1.0(\mathrm{syst})$ MeV/c$^2$
and width $\Gamma=17.5\pm5.2(\mathrm{stat})\pm5.9(\mathrm{syst})$ MeV by analyzing the $D^0p$ invariant mass spectrum, which is an isosinglet since there is no evidence of doubly charged partner in the $D^+p$ spectrum  \cite{Aubert:2006sp}. Later, the Belle Collaboration confirmed $\Lambda_c(2940)^+$ in $\Sigma_c(2455)^{0,++}\pi^{+,-}$ channels \cite{Abe:2006rz}, which gave $M=2938.0\pm1.3^{+2.0}_{-4.0}$ MeV/c$^2$ and $\Gamma=13^{+8+27}_{-5-7}$ MeV consistent with the BaBar's measurement \cite{Aubert:2006sp}.

The observation of $\Lambda_{c}(2940)^+$ has stimulated extensive interest
among different theoretical groups, which have proposed different explanations to the underlying structure of $\Lambda_c(2940)^+$. Since $\Lambda_c(2940)^+$ is near the threshold of $D^*p$, $\Lambda_c(2940)^+$ is explained as an S-wave $D^{*0}p$ molecular state with spin parity $J^{P}=\frac{1}{2}^-$, where the obtained decay behavior of $\Lambda_c(2940)^+$ is not only consistent with the experimental measurement but also is applied to test the molecular structure \cite{He:2006is}. Later, the strong decay of $\Lambda_c(2940)^+$ was studied under the $D^*N$ molecular state assignments with $J^{P}=\frac{1}{2}^-$ and $J^{P}=\frac{1}{2}^+$ in Ref. \cite{Dong:2009tg}, which indicates that $\Lambda_c(2940)^+$ should be assigned as the $D^*N$ molecular state with $J^{P}=\frac{1}{2}^+$. Recently the radiative decay of $\Lambda_c(2940)^+$ under the assignment of  the $D^*N$ molecular state with $J^{P}=\frac{1}{2}^+$ was performed in Ref. \cite{Dong:2010xv}. If $\Lambda_c(2940)^+$ is $J^P=\frac{1}{2}^+$ $D^*N$ molecular state, $D^*$ interacts with nucleon via P-wave.

Besides giving these exotic explanations to $\Lambda_{c}(2940)^+$, theorist has tried to find suitable assignment to $\Lambda_{c}(2940)^+$ under the framework of the conventional charmed baryon. The potential model once predicts the masses of $\Lambda_c$ with $J^{P}=\frac{5}{2}^-,\,\frac{3}{2}^+$ are 2900 MeV and 2910 MeV \cite{Capstick:1986bm,Copley:1979wj}, respectively. Cheng and Chua calculated the ratio of $\Sigma_c^*\pi/\Sigma_c\pi$ if $\Lambda_{c}(2940)^+$ is of $J^{P}=\frac{5}{2}^-$ or $\frac{3}{2}^+$ in heavy hadron chiral perturbation theory  \cite{Cheng:2006dk}, and indicated that such ratio is useful to distinguish the $J^P$ quantum number of $\Lambda_{c}(2940)^+$.
In Ref. \cite{Chen:2007xf}, the strong decays of the newly observed charmed hadrons have been calculated by the $^3P_0$ model. The corresponding numerical result indicates that $\Lambda_{c}(2940)^+$ could only be as D-wave charmed baryon $\check{\Lambda}_{c1}^0(\frac{1}{2}^+)$ or $\check{\Lambda}_{c1}^0(\frac{3}{2}^+)$ (see the notations for the D-wave charmed baryons in Ref. \cite{Chen:2007xf}) while $\Lambda_{c}(2940)^+$ as the first radial excitation of $\Lambda_c(2286)^+$ is fully excluded since $\Lambda_{c}(2940)^+\to D^0 p$ was observed by BaBar \cite{Aubert:2006sp}.
In the relativistic quark-diquark model, Ebert, Faustov and Galkin suggested $\Lambda_{c}(2940)^+$
as the first radial excitation of $\Sigma_c$ with $J^P=\frac{3}{2}^+$ \cite{Ebert:2007nw}.
The result obtained by chiral quark model indicates that $\Lambda_c(2940)^+$ is D-wave charmed baryon $\Lambda_c\,^2D_{\lambda\lambda}\frac{3}{2}^+$ \cite{Zhong:2007gp}. In Ref. \cite{Valcarce:2008dr}, $\Lambda_c(2940)^+$ as the first radial excitation of the $\Sigma_c$ with $J^P=\frac{3}{2}^+$ was proposed by solving the three-body problem by the Faddeev method in momentum space. By the mass load flux tube model, the authors in Ref. \cite{Chen:2009tm} suggested that $\Lambda_c(2940)^+$ could be as the orbitally excited $\Lambda_c^+$ with $J^P=\frac{5}{2}^-$.

\begin{center}
\begin{figure}[htb]
\begin{tabular}{c}
\scalebox{1}{\includegraphics{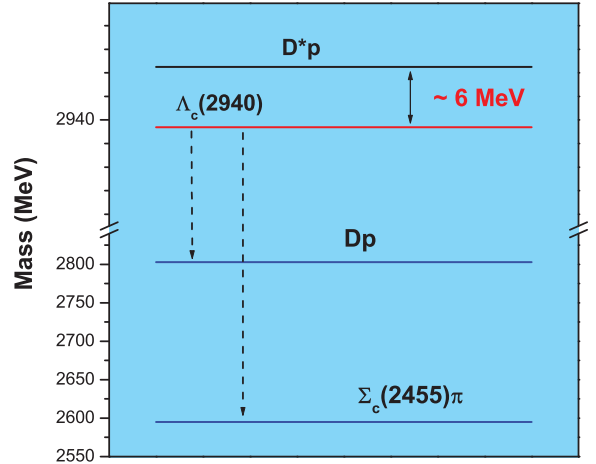}}
\end{tabular}
\caption{The observed decay modes of $\Lambda_c(2940)^+$ and the comparison of the mass of $\Lambda_c(2940)^+$
with the thresholds of $D^*p$, $Dp$, $\Sigma_c(2455)\pi$. \label{mass}}
\end{figure}
\end{center}

Although different theoretical explanations to $\Lambda_c(2940)^+$
were proposed, at present the properties of $\Lambda_c(2940)^+$ are still unclear, which means that more theoretical efforts are needed to reveal its underlying structure of $\Lambda_c(2940)^+$.

As shown in Fig. \ref{mass}, $\Lambda_c(2940)^+$ not only decays into $Dp$ and $\Sigma_c(2455)\pi$, but also is close to the threshold of $D^*p$, $i.e.$, about 6 MeV mass different between $\Lambda_c(2940)^+$ and the threshold of $D^*p$. Thus, exotic $D^*N$ molecular state becomes one of the possible explanations to the structure of $\Lambda_c(2940)^+$. The dynamical study of the $D^*N$ system in one-boson exchange model is an interesting research topic at present, which can help us further clarify the  $D^*N$ molecular state assignment to $\Lambda_c(2940)^+$ and deeply understand the $D^*N$ interaction. In this work, we systematically carry out the dynamical investigation of the $D^*N$ system.

This paper is organized as follows. After introduction, we present the detail of the dynamical study of the $D^*N$ system, which includes the relevant effective Lagrangian and coupling constants, the detailed derivation of the effective potential of $D^*N$ interaction, the corresponding numerical result, the study of $\bar{B}^*N$ system. Finally, the paper ends with the discussion and conclusion.

\section{The dynamical study of $D^*N$ system}\label{sec2}

\subsection{The effective Lagrangian and coupling constants}

In this work, we perform the dynamical study of $D^*N$ system. In order to deduce the effective
potential of the $D^*N$ interaction resulted from the pseudoscalar, vector and scalar meson
exchanges, we adopt effective Lagrangian approach. In this section, we
collect the relevant effective Lagrangian.

In terms of heavy quark limit and chiral symmetry, the Lagrangians depicting the interactions of
light pseudoscalar, vector and scalar mesons with S-wave
heavy flavor mesons were constructed in Refs.
\cite{Cheng1993,Yan1992,Wise1992,Burdman1992,Casalbuoni1997,Falk1992,Ding2009}
\begin{eqnarray}
\mathcal{L}_{HH\mathbb{P}}&=&
ig\langle H_b \gamma_\mu A_{ba}^\mu\gamma_5 \bar{H}_a\rangle
%
,\label{eq:lag}\\
\mathcal{L}_{HH\mathbb{V}}&=&
i\beta\langle H_b v_\mu
(\mathcal{V}^\mu_{ba}-\rho^\mu_{ba})\bar{H}_a\rangle
\nonumber\\&&
+i\lambda\langle H_b
\sigma_{\mu\nu}F^{\mu\nu}(\rho)\bar{H}_a\rangle
%
,\\
\mathcal{L}_{ HH\sigma}&=&g_s \langle H_a\sigma
\bar{H}_a\rangle
,
\label{eq:lag2}
\end{eqnarray}
where the multiplet field $H$ is composed of pseudoscalar
${\mathcal{P}}$ and vector ${\mathcal{P}}^{*}$ with ${\mathcal{P}}^{(*)T}
=(D^{(*)+},D^{(*)0})$ or
$(\bar{B}^{(*)0},B^{(*)-})$. And $H$ 
is defined by
\begin{eqnarray}
	H_a&=&\frac{1+\rlap\slash
	v}{2}[{\mathcal{P}}^{*}_{a\mu}\gamma^\mu
	-{\mathcal{P}}_a\gamma_5].
\end{eqnarray}
Here, $\bar{H}=\gamma_0H^\dag\gamma_0$ and $v=(1,{\mathbf 0})$.

In the above expressions, the $\mathcal{P}
$ and $\mathcal{P}^*
$ satisfy the normalization relations
$\langle 0|{\mathcal{P}}|Q\bar{q}(0^-)\rangle
=\sqrt{M_\mathcal{P}}$
and
$\langle
0|{\mathcal{P}}^*_\mu|Q\bar{q}(1^-)\rangle=
\epsilon_\mu\sqrt{M_{\mathcal{P}^*}}$. The axial
current is $A^\mu=\frac{1}{2}(\xi^\dag\partial_\mu\xi-\xi \partial_\mu
\xi^\dag)=\frac{i}{f_\pi}\partial_\mu{\mathbb P}+\cdots$ with
$\xi=\exp(i\mathbb{P}/f_\pi)$ and $f_\pi=132$ MeV.
$\rho^\mu_{ba}=ig_{V}\mathbb{V}^\mu_{ba}/\sqrt{2}$,
$F_{\mu\nu}(\rho)=\partial_\mu\rho_\nu - \partial_\nu\rho_\mu +
[\rho_\mu,{\ } \rho_\nu]$,
 and $g_V=m_\rho/f_\pi$. Here, $\mathbb P$ and $\mathbb V$ are two
by two pseudoscalar and vector matrices
\begin{eqnarray}
	{\mathbb P}=\left(\begin{array}{ccc}
		\frac{1}{\sqrt{2}}\pi^0&\pi^+\\
		\pi^-&-\frac{1}{\sqrt{2}}\pi^0
\end{array}\right),\qquad
\mathbb{V}=\left(\begin{array}{ccc}
\frac{\rho^{0}}{\sqrt{2}}+\frac{\omega}{\sqrt{2}}&\rho^{+}\\
\rho^{-}&-\frac{\rho^{0}}{\sqrt{2}}+\frac{\omega}{\sqrt{2}}
\end{array}\right).\nonumber
\end{eqnarray}

By expanding Eqs. (\ref{eq:lag})-(\ref{eq:lag2}), one further obtains the effective Lagrangian of
light pseudoscalar meson $\mathbb{P}$ coupling with heavy flavor mesons
\begin{eqnarray}\label{eq:lag-p-exch}
\mathcal{L}_{\mathcal{P}^*\mathcal{P}^*\mathbb{P}}
&=& -i\frac{2g}{f_\pi}\varepsilon_{\alpha\mu\beta\nu}
v^\alpha\mathcal{P}^{*\mu}_{b}{\mathcal{P}}^{*\nu\dag}_{a}
\partial^\beta{}\mathbb{P}_{ba}
,\label{ppp}\\
\mathcal{L}_{\mathcal{P}^*\mathcal{P}\mathbb{P}}
&=&- \frac{2g}{f_\pi}(\mathcal{P}^{}_b\mathcal{P}^{*\dag}_{a\lambda}+
\mathcal{P}^{*}_{b\lambda}\mathcal{P}^{\dag}_{a})\partial^\lambda{}\mathbb{P}_{ba}
.
\end{eqnarray}
The effective Lagrangian describing the coupling of light vector meson $\mathbb{V}$ and heavy flavor
mesons reads as
\begin{eqnarray}\label{eq:lag-v-exch}
  \mathcal{L}_{\mathcal{PP}\mathbb{V}}
  &=& -\sqrt{2}\beta{}g_V\mathcal{P}^{}_b\mathcal{P}_a^{\dag}
  v\cdot\mathbb{V}_{ba}
,\label{ppv}\\
  \mathcal{L}_{\mathcal{P}^*\mathcal{P}\mathbb{V}}
  &=&- 2\sqrt{2}\lambda{}g_V v^\lambda\varepsilon_{\lambda\alpha\beta\mu}
  (\mathcal{P}^{}_b\mathcal{P}^{*\mu\dag}_a +
  \mathcal{P}_b^{*\mu}\mathcal{P}^{\dag}_a)
  (\partial^\alpha{}\mathbb{V}^\beta)_{ba}
%
,\nonumber\\\label{ppsv}\\
  \mathcal{L}_{\mathcal{P}^*\mathcal{P}^*\mathbb{V}}
  &=& \sqrt{2}\beta{}g_V \mathcal{P}_b^{*}\mathcal{P}^{*\dag}_a
  v\cdot\mathbb{V}_{ba}\nonumber\\&&
  +i2\sqrt{2}\lambda{}g_V\mathcal{P}^{*\mu}_b\mathcal{P}^{*\nu\dag}_a
  (\partial_\mu{}
  \mathbb{V}_\nu - \partial_\nu{}\mathbb{V}_\mu)_{ba}
.\label{pspsv}
  \end{eqnarray}
The effective Lagrangian of the scalar $\sigma$ interacting with
heavy flavor mesons can be further expressed as
\begin{eqnarray}\label{eq:lag-s-exch}
  \mathcal{L}_{\mathcal{PP}\sigma}
  &=& -2g_s\mathcal{P}^{}_b\mathcal{P}^{\dag}_b\sigma
,\\
  \mathcal{L}_{\mathcal{P}^*\mathcal{P}^*\sigma}
  &=& 2g_s\mathcal{P}^{*}_b\cdot{}\mathcal{P}^{*\dag}_b\sigma
 .
\end{eqnarray}
As calculated in Ref.
~\cite{Isola2003}, $g=0.59$ in Eqs. (\ref{ppv})-(\ref{ppsv}) is obtained by the full width of $D^{*+}$ determined by experiment. The parameter $\beta$ appearing in the Lagrangians relevant to vector meson can be
fixed as $\beta=0.9$ by vector meson dominance mechanism while $\lambda=0.56$ GeV$^{-1}$ can be obtained by comparing the form factor calculated by light cone sum rule with that obtained by lattice QCD.
As the coupling constant related to scalar meson $\sigma$, $g_s=g_\pi/(2\sqrt{6})$ with $g_\pi=3.73$ was given in Refs. \cite{Falk1992,Liu2009}.

The effective vertices depicting the interaction of nucleon with
pseudoscalar meson $\mathbb{P}$, vector meson $\mathbb{V}$ and scalar
meson $\sigma$ are respectively
\begin{eqnarray}
\mathcal{L}_{{\mathbb P}NN}&=&-\frac{g_{ {\mathbb
P}NN}}{\sqrt{2}m_N} \bar{N}_b\gamma_5\gamma_\mu
\partial_\mu{\mathbb P}_{ba}  N_a,\\
\mathcal{L}_{\mathbb{V}NN}&=&-\sqrt{2}g_{\mathbb{V} NN}
\bar{N}_b\bigg(\gamma_\mu+\frac{\kappa}{2m_N}\sigma_{\mu\nu}\partial^\nu\bigg){\mathbb{V}}
_{ba}^\mu N_a,\label{vv}
\\
\mathcal{L}_{\sigma NN}&=&g_{\sigma NN} \bar{N}_a\sigma N_a,
\end{eqnarray}
where $N^T=(p,n)$ represents the nucleon field. The coupling
constants $g^2_{\pi NN}/(4\pi)=13.6$, $g^2_{\rho NN}/(4\pi)=0.84$,
$g^2_{\omega NN}/(4\pi)=20$ and $g^2_{\sigma NN}/(4\pi)=5.69$ with
$\kappa=6.1~(0)$ in Eq. (\ref{vv}) for $\rho~(\omega)$, which are
used in the Bonn nucleon-nucleon potential ~\cite{Machleidt2001d} and
meson productions in nucleon-nucelon
collision~\cite{Cao2010,Tsushima1998,Engel1996}. We follow the convention for
the signs of coupling constants as given in Refs.~\cite{Cao2010,Tsushima1998,Engel1996}.

We need to emphasize that in this work we only consider $\pi$, $\rho$,
$\omega$ and $\sigma$ exchanges due to the weak coupling of $\eta$ or
$\phi$ to nucleons as indicated in many previous
works \cite{Machleidt2001d,Cao2010}.

\subsection{Derivation of the effective potential of $D^*N$ interaction}

The scattering $D^*N\to
D^*N$ occurs via $\pi$, $\rho/\omega$ and $\sigma$ exchanges.

The scattering amplitude $i\mathcal{M}(J,J_Z)$, which is obtained by
effective Lagrangian approach, is related to the interaction potential
in the momentum space in terms of the Breit approximation
\begin{eqnarray}
V(\mathbf{q})=-\frac{1}{\sqrt{\prod_i 2M_i \prod_f
2M_f}}\mathcal{M}(J,J_Z) \; ,
\end{eqnarray}
where $M_{i}$ and $M_j$ denote the masses of the initial and final
states, respectively. The potential in the coordinate space
$V(\mathbf{r})$ is obtained after Fourier transformation.
For compensating the off-shell effect of exchanged particle and
describing the inner structure of every interaction vertex, the form
factor is introduced with monopole form
$F({\bm q}^2)=(\Lambda^2-m_i^2)/(\Lambda^2-q^2)$ when writing out scattering
amplitude, where the
cutoff $\Lambda$ should be around 1 GeV~\cite{Liu2009}.

With the above preparation, the $\pi$ exchange potential between
heavy flavor meson $D^*$ and nucleon in the momentum space is obtained
\begin{eqnarray}
V_\pi(\mathbf{q})&=&
-\frac{g_{\pi NN}\,g}{2\sqrt{2}f_\pi m_N}
({\mbox{\boldmath $\mathcal{T}$}} \cdot {\mbox{\boldmath $q$}})
({\mbox{\boldmath $\sigma$}}\cdot {\mbox{\boldmath $q$}})
P({\bm q}^2)F({\bm q}^2)
~~{\mbox{\boldmath $\tau$}}_{D^*}\cdot{\mbox{\boldmath
$\tau$}}_N.\nonumber\\\label{Eq:V}
\end{eqnarray}
The  $\rho$ exchange potential can be written as
\begin{eqnarray}
V_\rho(\mathbf{q})&=&
g_{\rho
NN}g_V\bigg\{-\beta\frac{1}{2}(\bm{\epsilon}^{m'\dag}\cdot\bm{\epsilon}^m)
\bigg[1+(1+2\kappa)\frac{i\bm{\sigma}\cdot{\bm q}\times{\bm
Q}}{4m_N^2}\bigg]\nonumber\\&&+\lambda\bigg[i\frac{1}{m_N}\bm{\mathcal{T}}
\cdot{\bm q}\times{\bm Q}+\frac{(1+\kappa)}{2m_N}
\bm{\mathcal{T}}\times{\bm q}\cdot{\bm \sigma}\times{\bm q}\bigg]\bigg\}\nonumber\\&&\times
P({\bm q}^2)F({\bm q}^2)
{\bm\tau}_{D^*}\cdot{\bm\tau}_N.\label{ha1}
\end{eqnarray}
The $\omega$ meson exchange potential can be easily obtained by replacing the
relevant coupling constants and the mass of exchanged light meson, and
removing the isospin factor ${\bm\tau}_{D^*}\cdot{\bm\tau}_N$ and setting $\kappa=0$ in Eq. (\ref{ha1}). The $\sigma$ exchange potential
reads as
\begin{eqnarray}
V_\mathbb{\sigma}(\mathbf{q})&=&
g_{\sigma NN}g_s{\bm\epsilon}^{m'\dag}\cdot{\bm\epsilon}^m
\bigg(1-\frac{{\bm\sigma}\cdot{\bm q}\times{\bm Q}}{4m_N^2}\bigg)
P({\bm q}^2)F({\bm q}^2).\nonumber\\\label{Eq:Vs}
\end{eqnarray}
In the above expressions of the obtained potentials,
$\mathcal{T}^{\alpha}=i\varepsilon^{0\alpha\beta\gamma}\epsilon_\beta^{m'\dag}
\epsilon_\gamma^{m}$ and $ P({\bm q}^2)=\frac{1}{{\mbox{\boldmath $q$}}^2+m_i^2}$.
The polarization vectors are defined as ${\bm
\epsilon}^{\pm}=\mp\frac{1}{\sqrt{2}}(1,\pm i,0)$, ${\bm
\epsilon}^0=(0,0,1)$.
Here, $m_i$ is the mass of the exchanged meson for $D^*N\to
D^*N$ transition.

In this work, we focus on the $D^{*}N$ systems with the total angular momentum $J\leqslant
\frac{5}{2}$, which are of positive or negative parity. Such $D^{*}N$ systems
can be categorized as twelve groups according to the quantum number
$I(J^{P})$ of system, $i.e.$, the systems with $0(\frac{1}{2}^\pm)$,
$1(\frac{1}{2}^\pm)$, $0(\frac{3}{2}^\pm)$, $1(\frac{3}{2}^\pm)$,
$0(\frac{5}{2}^\pm)$ and $1(\frac{5}{2}^\pm)$. Each of the $D^{*}N$ systems with
$J=\frac{1}{2}$ is composed of two states
\begin{eqnarray}
	\Big|I(\frac{1}{2}^-)\Big\rangle:&&\ \ 	\Big|^2S_{\frac{1}{2}}\Big\rangle,\ \ \ \
\Big|^4D_{\frac{1}{2}}\Big\rangle;\label{Eq:wf1}\\
	\Big|I(\frac{1}{2}^+)\Big\rangle:&&\ \ 	\Big|^4P_{\frac{1}{2}}\Big\rangle,\ \ \ \
|^2P_{\frac{1}{2}}\rangle.\label{EQwf2}
\end{eqnarray}
And each of the $D^{*}N$ systems with $J=\frac{3}{2},\,\frac{5}{2}$ is constructed by three states
\begin{eqnarray}
\Big|I(\frac{3}{2}^-)\Big\rangle:&&\ \ 	\Big|^4S_{\frac{3}{2}}\Big\rangle,\ \ \ \
	\Big|^2D_{\frac{3}{2}}\Big\rangle,\ \ \ \
	\Big|^4D_{\frac{3}{2}}\Big\rangle;\\
\Big|I(\frac{5}{2}^-)\Big\rangle:&&\ \ 	\Big|^2D_{\frac{5}{2}}\Big\rangle,\ \ \ \
	\Big|^4D_{\frac{5}{2}}\Big\rangle,\ \ \ \
	\Big|^4G_{\frac{5}{2}}\Big\rangle;\\
\Big|I(\frac{3}{2}^+)\Big\rangle:&&\ \ 	\Big|^2P_{\frac{3}{2}}\Big\rangle,\ \ \ \
	\Big|^4P_{\frac{3}{2}}\Big\rangle,\ \ \ \
	\Big|^4F_{\frac{3}{2}}\Big\rangle;\\
\Big|I(\frac{5}{2}^+)\Big\rangle:&&\ \ 	\Big|^4P_{\frac{5}{2}}\Big\rangle,\ \ \ \
	\Big|^2F_{\frac{5}{2}}\Big\rangle,\ \ \ \
	\Big|^4F_{\frac{5}{2}}\Big\rangle.
	\label{Eq:wf3}
\end{eqnarray}
Here, we use notation $^{2S+1}L_J$ to show the concrete information,
which includes total spin $S$, angular momentum $L$, total angular
momentum $J$ of the $D^{*}N$ system. $S$, $P$, $D$, $F$ and $G$ indicate
that the couplings between heavy flavor meson $D^*$ and nucleon occur
via $S$-wave, $P$-wave, $D$-wave, $F$-wave and $G$-wave interactions respectively, which means that in
this work we will include such $i$-wave contributions ($i=S,\,P,\,D,\,F,\,G$).

The general expressions of these states in Eqs.~(\ref{Eq:wf1})-(\ref{Eq:wf3})
can be explicitly written as
\begin{eqnarray}
	\Big|^{2S+1}L_J\Big\rangle	=\sum_{m,m',m_L,m_S}C_{Sm_S,Lm_L}^{JM}
	C_{\frac{1}{2}m,1m'}^{Sm_S}
	\epsilon^{m'}_{n}\chi_{\frac{1}{2}
	m}Y_{Lm_L},\label{po}
\end{eqnarray}
where $C_{\frac{1}{2}m,Lm_L}^{JM}$, $C_{Sm_S,Lm_L}^{JM}$ and $C_{\frac{1}{2}m,1m'}^{Sm_S}$ are Clebsch-Gordan coefficients.
$Y_{Lm_L}$ is spherical harmonics function. $\chi_{\frac{1}{2} m}$ denotes spin wave function of the corresponding state.
The polarization vector for $D^*$ is defined as $\epsilon^m_\pm=\mp\frac{1}{\sqrt{2}}(\epsilon^m_x\pm i\epsilon^m_y)$
and $\epsilon^m_0=\epsilon^m_z$. Here, the
polarization vector in Eq. (\ref{po}) is just the one appearing in the
potentials listed in Eqs.~(\ref{Eq:V})-(\ref{Eq:Vs}).

According to the sub-potentials in Eqs.~(\ref{Eq:V})-(\ref{Eq:Vs}) and
wave functions in Eqs.~(\ref{Eq:wf1})-(\ref{po}), one obtains the
total potential of the $D^{*}N$ system with
$J^P=\frac{1}{2}^\pm$,
\begin{widetext}
\begin{eqnarray}
&&V_{\frac{1}{2}^-} =	\left(\begin{array}{ccc}
-D-2C&\sqrt{2}T\\\sqrt{2}T&-D+C+18O'-O-2T
\end{array}\right),\\
&&V_{\frac{1}{2}^+} =	\left(\begin{array}{ccc}
-D-2C+8O'+\frac{2}{9}O&\sqrt{2}T+2\sqrt{2}O'+\frac{2\sqrt{2}}{9}O\\
\sqrt{2}T+2\sqrt{2}O'+\frac{2\sqrt{2}}{9}O&-D+C+10O'-\frac{5}{9}O-2T
\end{array}\right),
\label{33}
\end{eqnarray}
\end{widetext}
which are two by two matrixes since the $D^*N$ system with
$J^P=\frac{1}{2}^-$ or $J^P=\frac{1}{2}^+$ is constructed by two states just listed in
Eq.~(\ref{Eq:wf1}) or (\ref{EQwf2}).
Analogously, the total potential for the $D^*N$ system with
$J^P=\frac{3}{2}^\pm$ or $\frac{5}{2}^\pm$ can be expressed by three by
three matrix as
\begin{widetext}
\begin{eqnarray}
V_{\frac{3}{2}^-}&=&	\left(\begin{array}{cccc}
C &2T&-T\\
2T&C-\frac{8}{5}D-\frac{4}{3}O+12O'&\frac{2}{5}D+\frac{14}{15}O+6O'+T\\
-T&\frac{2}{5}D+\frac{14}{15}O+6O'+T&-D-2C+12O'+\frac{1}{15}O
	\end{array}\right),\\
V_{\frac{5}{2}^-}&=&	\left(\begin{array}{cccc}
-2C-D-8O'+\frac{2}{3}O&\frac{1}{3}\sqrt{\frac{2}{7}}(-6D+42O'+4O-3T)
&2\sqrt{\frac{3}{7}}T\\
\frac{1}{3}\sqrt{\frac{2}{7}}(-6D+42O'+4O-3T)
&C-\frac{2}{21}(-12D+8O+21O'+15T)&\frac{4\sqrt{6}}{7}T\\
2\sqrt{\frac{3}{7}}T&\frac{4\sqrt{6}}{7}T
&C+\frac{1}{14}[-9D+5(84O'-3O-4T)]
	\end{array}\right),\nonumber\\ \\
V_{\frac{3}{2}^+}&=&	\left(\begin{array}{cccc}
-2C-D-4O'+\frac{1}{3}O&\frac{1}{3\sqrt{5}}(-6D+3O'+2O-3T)
&\frac{3}{\sqrt{5}}T\\
\frac{1}{3\sqrt{5}}(-6D+3O'+2O-3T)
&C-\frac{4}{15}(-3D+O+15O'+6T)&\frac{6}{5}T\\
\frac{3}{\sqrt{5}}T&\frac{6}{5}T
&C-\frac{4}{15}(3D-90O'+4O+6T)
	\end{array}\right),\\
V_{\frac{5}{2}^+}&=&	\left(\begin{array}{cccc}
	C-6D-4O'+\frac{2}{5}T&-\sqrt{\frac{6}{5}}T&\frac{4\sqrt{6}}{5}T\\
	-\sqrt{\frac{6}{5}}T&-2C-D+16O'-\frac{4}{21}O
	&\frac{2}{21\sqrt{5}}[15D+7(30O'+5O+3T)]\\
	\frac{4\sqrt{6}}{5}T&\frac{2}{21\sqrt{5}}[15D+7(30O'+5O+3T)]
	&C-\frac{25}{14}D+14O'-\frac{85}{42}O+\frac{2}{5}6T
	\end{array}\right)
	\label{Eq: V}
\end{eqnarray}
\end{widetext}
with
\begin{eqnarray}
	D&=&D'^\beta_\rho{\bm \tau}_{D^*}\cdot{\bm \tau}_N
	+D'^\beta_\omega-D'_\sigma,\\
	C&=&-C_\pi- 2[(1+\kappa)C_\rho^\lambda{\bm
	\tau}_{D^*}\cdot{\bm
	\tau}_N+C_\omega^\lambda],\label{Eq:CDT-1}\\
	T&=&-T_\pi+ 2[(1+\kappa)T_\rho^\lambda{\bm
	\tau}_{D^*}\cdot{\bm
	\tau}_N+T_\omega^\lambda],\label{Eq:CDT-2}\\
	O&=&O_\sigma-[(1+2\kappa)O_\rho^\beta{\bm \tau}_{D^*}
	\cdot{\bm\tau}_N+O_\omega^\beta],\\
	O'&=&O_\rho^\lambda{\bm \tau}_{D^*}\cdot{\bm
	\tau}_N+O_\omega^\lambda,
\end{eqnarray}
where the expressions of $C_i$, $D'_i$, $T_i$ and $O_i$ ($i=\pi,\,\sigma$) are defined as
\begin{eqnarray}
D'_i&=&4m_N^2D_i\nonumber\\&=&4m_N^2\bigg\{\frac{1}{4\pi}~F_i\bigg[Y_0(m_i,r)-Y_0(\Lambda,r)
\nonumber\\&&-\frac{\xi_i^2}{2\Lambda}rY_0(\Lambda,
r)\bigg]\bigg\},\label{11}
\end{eqnarray}
\begin{eqnarray}
C_i&=&\frac{1}{4\pi}~F_i\bigg[Z_0(m_i,r)Y_0(m_i, r)
-Z_0(\Lambda,r)Y_0(\Lambda, r)
\nonumber\\&&-\frac{\xi_i^2}{2}(\Lambda r-2)Y_0(\Lambda, r)\bigg],\\
T_i&=&\frac{1}{4\pi}~F_i\bigg[T_0(m_i,r)Y_0(m_i, r)-T_0(\Lambda,r)Y_0(\Lambda,r)
\nonumber\\&&-\frac{\xi_i^2}{2}(\Lambda r+1)Y_0(\Lambda,r)\bigg],\\
O_i&=&\frac{1}{4\pi}~F_i\bigg[O_0(m_i,r)Y_0(m_i, r)-O_0(\Lambda,r)Y_0(\Lambda,r)
\nonumber\\&&-\frac{\xi_i^2}{2}rY_0(\Lambda,r)\bigg]\label{Eq:CDT}.
\end{eqnarray}
The expressions of $C_i^j$, $D_i^{\prime j}$, $T_i^j$ and $O_i^j$ ($i=\rho,\,\omega$
and $j=\beta,\lambda$) are similar to those of $C_i$, $D'_i$, $T_i$ and $O_i$
respectively, which can be easily obtained by replacing
factor $F_i$ with $F_i^j$ in Eqs. (\ref{11})-(\ref{Eq:CDT})
($F_\pi=g g_{\pi NN}/(6\sqrt{2}m_Nf_\pi)$,
$F_V^\beta=g_{V NN}g_V\beta/(4m_N^2)$, $F_V^\lambda=g_{V NN}g_V\lambda/(3m_N)$,
$F_\sigma=g_{\sigma NN}g_\sigma/(4m_N^2)$). Here,
$\xi_i=\sqrt{\Lambda^2-m_i^2}$ and
functions $Y_0(x,r)$, $Z_0(x,r)$, $T_0(x,r)$ and $O_0(x,r)$ are
\begin{eqnarray}
Y_0(x,r)&=&\frac{e^{-xr}}{r}, \qquad Z_0(x,r)=x^2,\nonumber\\
T_0(x,r)&=&x^2\Big[1+\frac{3}{xr}+\frac{3}{(xr)^2}\Big],\nonumber\\ O_0(x,r)&=&x^2\Big[\frac{1}{xr}+\frac{1}{(xr)^2}\Big].\nonumber
\end{eqnarray}

\subsection{Numerical result}
As shown in Sec. \ref{sec2}, the positive values of parameters $g$,
$\beta/\lambda$ and $g_s$
\cite{Casalbuoni1997,Falk1992,Isola2003,Liu2009}, which are relevant
to pseudoscalar, vector and scalar meson exchanges respectively, are
widely adopted in previous theoretical work. Thus, we first present the numerical result under taking the positive values of parameters $g$,
$\beta/\lambda$ and $g_s$.

\begin{figure}[h!]
\begin{center}
\includegraphics[bb=10 30 540 400 ,scale=0.45]{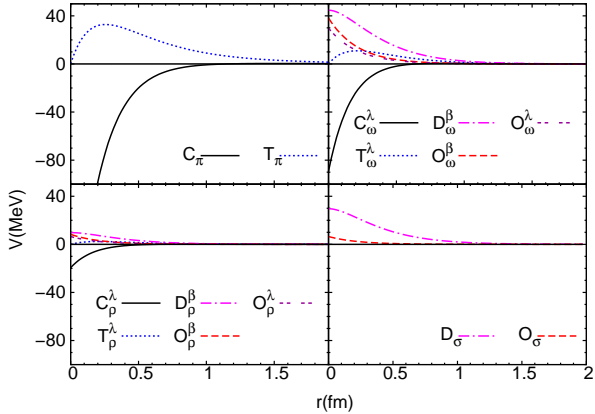}
\caption{(Color online). The dependence of $C_i$, $D_i$,
$T_i$, $O_i$, $C_i^j$, $D_i^j$,
$T_i^j$, $O_i^j$ on $r$ for the $D^{*}N$ system. Here, we use the $D^{*}N$ system with isospin $I=1$ as an example.
The cutoff $\Lambda=1$ GeV and the signs of $g$, $\beta/\lambda$, $g_s$ are taken as positive.
\label{Fig:V}}
\end{center}
\end{figure}

With the isovector $D^*N$ system as an example, one shows the line shapes of $C_i$, $D_i$,
$T_i$, $O_i$, $C_i^j$, $D_i^j$, $T_i^j$, $O_i^j$ in Fig.~\ref{Fig:V}
under taking cutoff $\Lambda=1$ GeV, where the signs of $g$,
$\beta/\lambda$, $g_s$ are taken as positive. The pion exchange really
provides important contribution to the exchange potential as shown in
the left figure while the vector meson exchange also give considerable
contributions to the effective potential. The contribution from
$\omega$ meson exchange is comparable with that from pion exchange due
to the large coupling constant relevant to $\omega$. Compared with
$\pi$, $\rho$ and $\omega$ exchanges, scalar meson exchange only gives
small contribution to the effective potential. The spin-orbit terms
$O_i$ from the vector and scalar meson exchanges, which correspond to
the relativistic correction, are neglectable compared with the other terms.

Using the potential obtained above, the binding energy for the
$D^{*}N$ systems with
$J^P=\frac{1}{2}^\pm,\frac{3}{2}^\pm,\frac{5}{2}^\pm$ can be obtained
by solving the coupled-channel Schr\"odinger equation. One uses the
FESSDE program \cite{Abrashkevich1995} to produce the numerical
results.

\begin{figure}[h!]
\includegraphics[bb=180 120 530 370 ,scale=0.69]{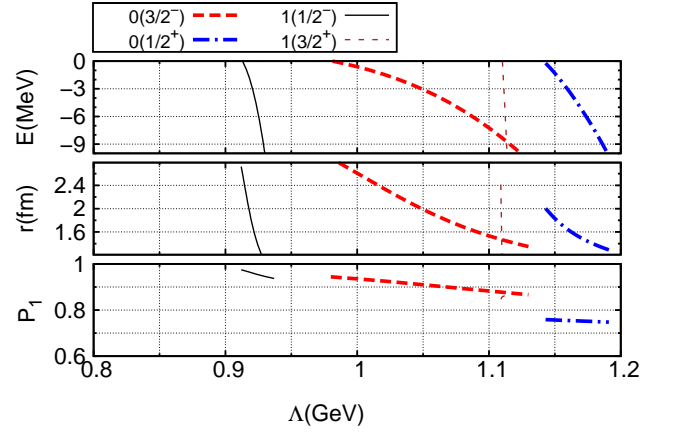}
\caption{
(Color online.) The $\Lambda$ dependence of the binding energy, the root-mean-square radius $r$
and $P_1$ which is the possibility of the $|^2S_{\frac{1}{2}}\rangle$, $|^4P_{\frac{1}{2}}\rangle$, $|^4S_{\frac{3}{2}}\rangle$ and $|^2P_{\frac{3}{2}}\rangle$ states in the ${D}^*N$ systems with
$\frac{1}{2}^-$, $\frac{1}{2}^+$, $\frac{3}{2}^-$ and
$\frac{3}{2}^+$, respectively.
The thick dashed, the thick dash-dotted, the solid and the dashed lines are
the bound state solutions of the $D^*N$ systems with $I(J^P)=0(\frac{3}{2}^-),
0(\frac{1}{2}^+), 1(\frac{1}{2}^-), 1(\frac{3}{2}^+)$ respectively.
\label{Fig:ED}}
\end{figure}

The obtained binding energy and the relevant root-mean-square radius
$r$ (in the unit of fm) of the $D^{*}N$ systems are presented in
Fig.~\ref{Fig:ED} with the
variation of the cutoff $\Lambda$ in the region of $0.8\leq\Lambda\leq1.2$ GeV.
Here, we only show the bound state solution with binding energy less
than 10 MeV since the OBE model is valid to deal with the loosely
bound hadronic molecular system. As shown in Fig.~\ref{Fig:ED}, we can
find bound state solutions only for four $D^*N$ systems with
$I(J^P)=0(\frac{3}{2}^-),\,0(\frac{1}{2}^+),\,1(\frac{1}{2}^-)
,\,1(\frac{3}{2}^+)$ among twelve systems shown in
Eqs.~(\ref{Eq:wf1})-(\ref{Eq:wf3}). In Fig.~\ref{Fig:ED}, one also presents the possibilities of $|^2S_{\frac{1}{2}}\rangle$, $|^4P_{\frac{1}{2}}\rangle$, $|^4S_{\frac{3}{2}}\rangle$ and $|^2P_{\frac{3}{2}}\rangle$ components
appearing in the corresponding ${D}^*N$ systems with
$\frac{1}{2}^-$, $\frac{1}{2}^+$, $\frac{3}{2}^-$ and
$\frac{3}{2}^+$, which indicates that $|^2S_{\frac{1}{2}}\rangle$, $|^4P_{\frac{1}{2}}\rangle$, $|^4S_{\frac{3}{2}}\rangle$ and $|^2P_{\frac{3}{2}}\rangle$ states are dominant in the $D^*N$ systems with
$\frac{1}{2}^-$, $\frac{1}{2}^+$, $\frac{3}{2}^-$ and $\frac{3}{2}^+$ respectively.

\begin{figure}[h!]
\includegraphics[bb=35 420 470 740,scale=0.57]{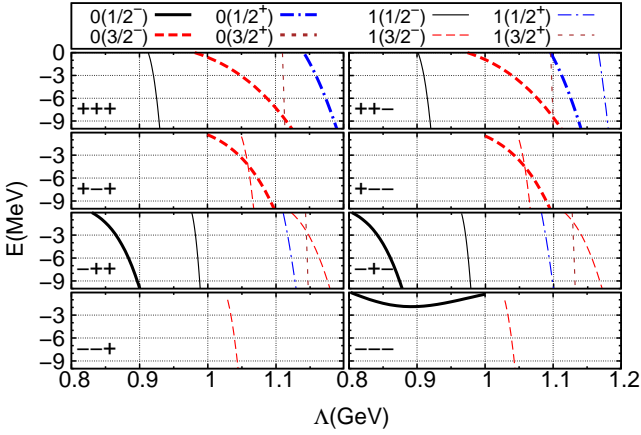}
\caption{(Color online.)
The binding energy for the $D^*N$ systems. The solid line, the dashed line, the
dash-dotted line, and the dotted lines are for the bound state
solutions with $J^P=\frac{1}{2}^-, \frac{3}{2}^-, \frac{1}{2}^+,
\frac{3}{2}^+$ respectively. Here, $+/-$ in $``\pm1\pm1\pm1"$ denotes that we need to multiply
corresponding pion, vector and sigma exchange potentials of the $D^{*}N$
systems listed in Eqs. (\ref{Eq:V})-(\ref{Eq:Vs}) by an extra factor
$+1/-1$, which come from the changes of the signs of coupling
constants.
\label{Fig:Dsign}}
\end{figure}

As an isoscalar state, $\Lambda^+_c(2940)$ can directly correspond to one of
the $D^*N$ systems with $I(J^P)=0(\frac{1}{2}^\pm)$, $0(\frac{3}{2}^\pm)$,
$0(\frac{5}{2}^\pm)$.  If $\Lambda^+_c(2940)$ is $D^*N$ molecular
state, such $D^*N$ molecular
state should be of the binding energy $-6$ MeV. Among the above six
possible $D^*N$ systems, only two $D^*N$ systems with $I(J^P)=0(\frac{1}{2}^+)$, $0(\frac{3}{2}^-)$ are of $-6$ MeV binding energy under taking cutoff $\Lambda$ as $1.17$ GeV and $1.09$ GeV, respectively. Thus, our dynamics study presented in this work
support to explain $\Lambda^+_c(2940)$ as $D^*N$ system with
$I(J^P)=0(\frac{1}{2}^+)$ or $0(\frac{3}{2}^-)$.

Although positive values for parameters $g$, $\beta/\lambda$, $g_s$
are adopted in former work
\cite{Casalbuoni1997,Falk1992,Isola2003,Liu2009} corresponding to the
case with $``+++"$ in Fig. \ref{Fig:Dsign}, in fact the signs of these parameters can not be
well constrained by the experiment data or theoretical calculation,
which could results in changing the signs of corresponding pion,
vector and sigma exchange potentials of $D^{*}N$ systems. As shown in
Fig.~\ref{Fig:Dsign}, one presents the binding energy dependent on
$\Lambda$ under eight combinations of the signs of $g$,
$\beta/\lambda$, $g_s$, where $+/-$ denotes that we need to multiply
corresponding pion, vector and sigma exchange potentials of $D^{*}N$
systems listed in Eqs. (\ref{Eq:V})-(\ref{Eq:Vs}) by an extra factor
$+1/-1$ which results from the changes of the signs of $g$,
$\beta/\lambda$, $g_s$.

We find that the sigma exchange contribution can be negligible since
the line shapes of binding energy dependent on $\Lambda$ shown in the
second column almost keep the same as those in the first column as
describing in Fig.~\ref{Fig:Dsign}. $\pi$ and $\rho/\omega$ meson
exchanges play very important role to form the $D^*N$ bound state.
These observations are consistent with the expected behavior of the potential of $D^*N$.

Our numerical results shown in Fig.~\ref{Fig:Dsign} indicate that there do not exist $D^*N$ molecular states with $J^P=\frac{5}{2}^\pm$.


\subsection{Bottom partner of the $D^*N$ system}

\begin{figure}[h!]
\includegraphics[bb=180 120 530 370,scale=0.7]{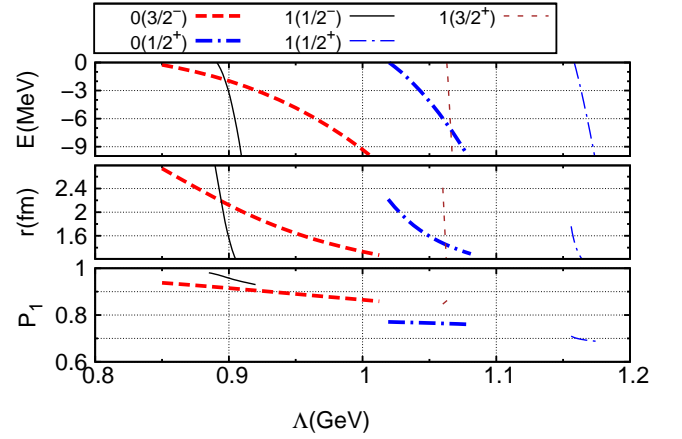}
\caption{
(Color online.) The $\Lambda$ dependence of the binding energy, the root-mean-square radius $r$
and $P_1$ which is the possibility of the $|^2S_{\frac{1}{2}}\rangle$, $|^4P_{\frac{1}{2}}\rangle$, $|^4S_{\frac{3}{2}}\rangle$ and $|^2P_{\frac{3}{2}}\rangle$ states in the $\bar{B}^*N$ systems with
$\frac{1}{2}^-$, $\frac{1}{2}^+$, $\frac{3}{2}^-$ and
$\frac{3}{2}^+$, respectively.
The thick dashed, the thick dash-dotted, the solid and the dashed lines are the bound state solutions of the $\bar{B}^*N$ systems with $I(J^P)=0(\frac{3}{2}^-),
0(\frac{1}{2}^+), 1(\frac{1}{2}^-), 1(\frac{1}{2}^+), 1(\frac{3}{2}^+)$ respectively.
\label{Fig:EB}}
\end{figure}

\begin{figure}[htb!]
\includegraphics[bb=35 420 470 740,scale=0.56]{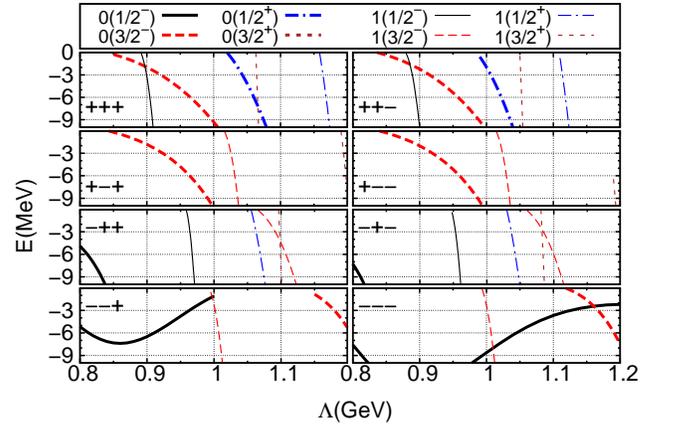}
\caption{
The binding energy for the $\bar{B}^*N$ systems. The solid line, the dashed line, the
dash-dotted line, and the dotted lines are for the bound state
solutions with $J^P=\frac{1}{2}^-, \frac{3}{2}^-, \frac{1}{2}^+,
\frac{3}{2}^+$ respectively. Here, $+/-$ in $(\pm1\pm1\pm1)$ denotes that we need to multiply
corresponding pion, vector and sigma exchange potentials of $\bar{B}^{*}N$
systems listed in Eqs. (\ref{Eq:V})-(\ref{Eq:Vs}) by an extra factor
$+1/-1$, which come from the changes of the signs of coupling
constants.
\label{Fig:Bsign}}
\end{figure}

In the previous section, the $D^*N$ system are investigated and the
bound states solutions are found. The observed $\Lambda_c(2940)^+$ can
be assigned as the $D^*N$ molecular state with $0(\frac{1}{2}^+)$ or
$0(\frac{3}{2}^-)$ supported by the dynamics study of the $D^*N$ system.
Due to the heavy quark symmetry, we also extend the same formulism to the
$\bar{B}^*N$ system, which is the bottom partner of the the $D^*N$ system.
The numerical results for the $\bar{B}^*N$ system are listed in
Figs.~\ref{Fig:EB} and \ref{Fig:Bsign}, which are similar to the
discussion for the $D^*N$ system.

If $\Lambda_c(2940)^+$ as $D^*N$ molecular state with
$I(J^P)=0(\frac{1}{2}^+)$ or $0(\frac{3}{2}^-)$, one expects that
there should exist the corresponding bottom partner of
$\Lambda_c(2940)^+$. As indicated in Fig.~\ref{Fig:EB}, one indeed
finds the bound state solutions of $\bar{B}^*N$ systems with
$0(\frac{1}{2}^+)$ and $0(\frac{3}{2}^-)$. Thus, the experimental
search for $\bar{B}^*N$ molecular state will be helpful to deep our
understanding of the underlying structure of $\Lambda_c(2940)^+$,
which might be as the task in LHCb.

\section{Conclusion and discussion}\label{sec5}

Stimulated by the observation of $\Lambda_c(2940)^+$
\cite{Aubert:2006sp,Abe:2006rz}, which is close to the threshold of
$D^*p$, we study the interaction of $D^*$ meson with nucleon $N$,
where the OBE model is applied to obtain the effective potential of
$D^*N$ system. By solving Schr\"{o}dinger equation, one can find the
bound state solution of the $D^*N$ system, which will be helpful to answer
whether there exists the $D^*N$ bound state corresponding to
$\Lambda_c(2940)^+$.

As indicated by the obtained numerical result, there exist bound state
solutions ($-6$ MeV binding energy) for the $D^*N$ system with
$I(J^P)=0(\frac{1}{2}^+)$, $0(\frac{3}{2}^-)$ taking reasonable
cutoff $\Lambda$, which indicate that it is possible to explain
$\Lambda_c(2940)^+$ as isoscalar S-wave or isoscalar P-wave
$D^*N$ molecular by performing the dynamical study. Additionally, we
find the bound state solutions for the isovector $D^*N$ systems.
Searching isovector $D^*N$ states might be as the task in future
experiment. In this work, we also predicted the existence of the bottom partners of
the $D^*N$ systems as the extension of the study of the $D^*N$ system. Carrying out the experimental search for the $\bar{B}^*N$ molecular states will be an interesting topic, especially for
LHCb experiment.

Searching for exotic nuclei is a very important research topic in
hadron physics and nuclear physics, which not only helps us to
understand the interaction of meson or hyperon with nucleon but also
provides the important information to reveal some underlying problems
in astrophysics. There are extensive studies of hypernucleus
\cite{Danysz1953,Pniewski1962,Danysz1963,Danysz1963a}, $\eta$-mesic
nucleus \cite{Haider1986a,Bhalerao1985a}. It is natural to expect the
exotic nuclei composed of vector heavy flavor meson and nucleon.
Experimental search for exotic nucleus composed of a vector heavy
flavor meson ($\bar{Q}q$ or $Q\bar{q}$ meson with $Q=c$ or $b$) and
the nucleon might be as the main task at J-PARC, RHIC and
FAIR~\cite{Riedl2007,Lutz2009} since the heavy flavor meson
($\bar{Q}q$ or $Q\bar{q}$ meson with $Q=c$ or $b$) can be produced in
a nucleon-rich environment, which also provides another approach
different from the production process of $\Lambda_c(2940)^+$ to test
the our prediction of the $D^*N/\bar{B}^*N$ molecular states to some
extent.

\section*{Acknowledgements}

We would like to thank Prof. Xin-Heng Guo, Dr. Dian-Yong Chen and Dr.
Yan-Rui Liu for fruitful discussions.  This project is supported by
the National Natural Sci- ence Foundation of China under Grants No.
10705001, No. 10905077, No. 11035006; the Foundation for the Author of National
Excellent Doctoral Dissertation of P.R. China (FANEDD) under Contracts
No. 200924; the Doctoral Program Foundation of Institutions of Higher
Education of P.R. China under Grant No. 20090211120029; the Program
for New Century Excellent Talents in University (NCET) by Ministry of
Education of P.R. China under Grant No. NCET-10-0442; the Fundamental
Research Funds for the Central Universities under Grant No. lzujbky-2010-69; the project sponsored by
SRF for ROCS, SEM under Grant No. HGJO90402; the Special Foundation of
President Support by the Chinese Academy of Sciences under Grant No.
YZ080425.

\end{document}